# Geometrically nonlinear isogeometric analysis of laminated composite plates based on higher-order shear deformation theory

Loc V. Tran[1,2], Jaehong Lee[2*], H. Nguyen-Van[3], H. Nguyen-Xuan[2,4], M. Abdel Wahab[1]

[1]Department of Mechanical Construction and Production, Faculty of Engineering and Architecture, Ghent University, 9000, Ghent – Belgium

[2]Department of Architectural Engineering, Sejong Unviersity, 98 Kunja Dong, Kwangjin Ku, Seoul, 143-747, South Korea

[3]Faculty of Civil Engineering, Ho Chi Minh City University of Architecture, 196 Pasteur Street, District 3, Ho Chi Minh City, Viet Nam

[4]Department of Computational Engineering, Vietnamese-German University, Binh Duong New City, Vietnam

**Abstract**

In this paper, we present an effectively numerical approach based on isogeometric analysis (IGA) and higher-order shear deformation theory (HSDT) for geometrically nonlinear analysis of laminated composite plates. The HSDT allows us to approximate displacement field that ensures by itself the realistic shear strain energy part without shear correction factors. IGA utilizing basis functions namely B-splines or non-uniform rational B-splines (NURBS) enables to satisfy easily the stringent continuity requirement of the HSDT model without any additional variables. The nonlinearity of the plates is formed in the total Lagrange approach based on the von-Karman strain assumptions. Numerous numerical validations for the isotropic, orthotropic, cross-ply and angle-ply laminated plates are provided to demonstrate the effectiveness of the proposed method.

---

[*] Corresponding author *Email address*: jhlee@sejong.ac.kr (Jaehong Lee)

# 1. Introduction

Various plate theories have been addressed since long time. Pagano [1] initially investigated the analytical three-dimensional (3D) elasticity method to predict the exact solution of simple static problems. Noor [2] had further developed the 3D elasticity solution formulas for stress analysis of composite structures. It is well known that the exact 3D approach is a potential tool to obtain the true solutions of plates. Nevertheless, it is not often easy to solve the practical problems that consider the complex (or even slightly complicated) geometries, arbitrary boundary conditions, and lamination schemes or nonlinearities. In addition, each layer in the 3D elasticity theory is modeled as a 3D solid so that the computational cost of laminated composite plate analyses is increased significantly. Hence, many equivalent single layer (ESL) plate theories with suitable assumptions [3] have been then proposed to transform the 3D problems to 2D problems.

Among the ESL plate theories, the classical laminate plate theory (CLPT) based on the Love – Kirchhoff assumption was proposed. However, the laminated plates stacking from many laminae are very susceptible to the transverse shear deformation effect due to the significantly smaller effective transverse shear modulus as compared to the effective elastic modulus along the fiber direction [4]. Thus, first order shear deformation theory (FSDT) based on Reissner– Mindlin theory [5, 6], which takes into account the shear effect, was therefore developed. In the FSDT model, with the linear in-plane displacement through plate thickness assumption, the obtained shear strains/stresses distribute inaccurately and do not satisfy the traction free boundary conditions at the plate surfaces. It is hence required to amend the unrealistic shear strain energy part by the shear correction factors (SCF). To overcome the limitation of the FSDT, various kinds of higher order shear deformable theory (HSDT) have then been devised (see Ref. [7] for a brief review). The HSDT models, which include higher-order terms in the approximation of the displacement field, ensure non-linear distributions of the shear strains/stresses with traction-free boundary condition at the plate surfaces. As a result, the HSDT models provide better results and yield more accurate and stable solutions (e.g. inter-laminar stresses and displacements) [8, 9] than the FSDT ones without the SCF. The HSDT requires the $C^1$-continuity of generalized displacement field leading to the second-order derivative of the stiffness formulation. However, the enforcement of even $C^1$ continuity across the inter-element boundaries in standard finite element method is not a trivial task. In attempts to overcome this difficulty, some kinds of methods have been developed such as $C^0$ continuous elements [10-12], nonconforming plate bending elements based on Hermite polynomial functions [3] or mixed finite elements [4,13]. It is known that there exists an algebraically complicated requirement in the construction of these elements. Furthermore, some extra unknown variables are needed to form their formulations which require much storage and computational cost. This shortcoming motivates us to develop in this paper a novel computational approach based on isogeometric analysis (IGA).

Isogeometric approach [14] firstly proposed by Hughes fulfills a seamless bridge link between computer aided design (CAD) and finite element analysis (FEA). IGA uses same B-Spline or non-uniform rational B-Spline (NURBS) functions in describing the exact geometry of problem and constructing finite approximation for analysis. Being thankful to higher order continuity of NURBS, IGA naturally verifies the $C^1$-continuity of plates based on the HSDT assumptions. IGA has been widely applied to the plate structures with various plate models such as CLPT [15], FSDT [16, 17], HSDT [7, 18-20], four unknown variables refined plate theory (RPT) [21, 22], layerwise [23, 24], etc. The literatures mentioned above, however, did not take into account geometric nonlinearity. So far, there are very few published materials related to geometrically nonlinear plate models using IGA, except two recent papers [25, 26] based on the FSDT. Apparently, there are no researches on geometrically nonlinear isogeometric analysis for the plates based on the HSDT model. Therefore, our goal in this paper is for the first time using HSDT model in study both geometrically nonlinear bending and transient analysis of the laminated composite plates. Based on the von-Karman strain which considers small strain and moderate rotation assumptions, the nonlinearity of the plates is formulated using total Lagrange approach and solved by the Newmark time integration association with the iteration methods. Several numerical examples are given to show the effectiveness of the present formulation in comparison with other available procedures in the literature.

The paper is outlined as follows. The next section introduces the generalized higher-order shear deformation theory for laminated composite plate. In section 3, the geometrically nonlinear formulations of plate based on IGA are described. Section 4 presents the solution scheme to solve the nonlinear problems. The numerical results and discussions are provided in section 5. Finally, this article is closed with some concluding remarks.

## 2. Generalized higher-order shear deformation theory for laminated composite plate

According to a generalized higher-order shear deformation theory [8], the displacements of an arbitrary point in the plate can be expressed in the general form

$$\mathbf{u} = \mathbf{u}_1 + z\mathbf{u}_2 + f(z)\mathbf{u}_3 \qquad (1)$$

where $\mathbf{u}_1 = \{u_0 \ v_0 \ w\}^T$ is the axial displacement, $\mathbf{u}_2 = -\{w_{,x} \ w_{,y} \ 0\}^T$ and $\mathbf{u}_3 = \{\beta_x \ \beta_y \ 0\}^T$ are the rotations in the $x$, $y$ and $z$ axes, respectively. The function $f(z)$ is the so-called distributed function which is set $f(z) = z - \frac{4}{3h^2}z^3$ follow to the famous Reddy's plate theory [27].

For a bending plate, the Green strain vector is expressed by

$$\varepsilon_{ij} = \frac{1}{2}\left(\frac{\partial u_i}{\partial x_j} + \frac{\partial u_j}{\partial x_i}\right) + \frac{1}{2}\frac{\partial u_k}{\partial x_i}\frac{\partial u_k}{\partial x_j} \qquad (2)$$

Using the von-Karman assumptions, the nonlinear strain – displacement relation adopts here by neglecting second-order terms of $u_0$ and $v_0$ displacements

$$\begin{Bmatrix}\boldsymbol{\varepsilon}\\ \boldsymbol{\gamma}\end{Bmatrix} = \begin{Bmatrix}\boldsymbol{\varepsilon}_m\\ 0\end{Bmatrix} + \begin{Bmatrix}z\boldsymbol{\kappa}_1\\ 0\end{Bmatrix} + \begin{Bmatrix}f(z)\boldsymbol{\kappa}_2\\ f'(z)\boldsymbol{\beta}\end{Bmatrix} \qquad (3)$$

where

$$\boldsymbol{\varepsilon}_m = \begin{bmatrix}u_{0,x}\\ v_{0,y}\\ u_{0,y}+v_{0,x}\end{bmatrix} + \frac{1}{2}\begin{bmatrix}w_{,x}^2\\ w_{,x}^2\\ 2w_{,xy}\end{bmatrix} = \boldsymbol{\varepsilon}_L + \boldsymbol{\varepsilon}_{NL}$$

$$\boldsymbol{\kappa}_1 = -\begin{bmatrix}w_{,xx}\\ w_{,yy}\\ 2w_{,xy}\end{bmatrix}, \quad \boldsymbol{\kappa}_2 = \begin{bmatrix}\beta_{x,x}\\ \beta_{y,y}\\ \beta_{x,y}+\beta_{y,x}\end{bmatrix}, \quad \boldsymbol{\beta} = \begin{bmatrix}\beta_x\\ \beta_y\end{bmatrix} \qquad (4)$$

and the nonlinear component of in-plane strain can be rewritten as

$$\varepsilon_{NL} = \frac{1}{2}\mathbf{A}_\theta \boldsymbol{\theta}$$

where $\quad \mathbf{A}_\theta = \begin{bmatrix} w_{,x} & 0 \\ 0 & w_{,y} \\ w_{,y} & w_{,x} \end{bmatrix}$ and $\boldsymbol{\theta} = \begin{Bmatrix} w_{,x} \\ w_{,y} \end{Bmatrix}$ (5)

For a $k^{th}$ orthotropic layer, the constitutive equation in local coordinate is derived from Hooke's law:

$$\begin{Bmatrix} \sigma_{xx} \\ \sigma_{yy} \\ \tau_{xy} \\ \tau_{xz} \\ \tau_{yz} \end{Bmatrix}^{(k)} = \begin{bmatrix} Q_{11} & Q_{12} & Q_{16} & 0 & 0 \\ Q_{21} & Q_{22} & Q_{26} & 0 & 0 \\ Q_{61} & Q_{62} & Q_{66} & 0 & 0 \\ 0 & 0 & 0 & Q_{55} & Q_{54} \\ 0 & 0 & 0 & Q_{45} & Q_{44} \end{bmatrix}^{(k)} \begin{Bmatrix} \varepsilon_{xx} \\ \varepsilon_{yy} \\ \gamma_{xy} \\ \gamma_{xz} \\ \gamma_{yz} \end{Bmatrix}^{(k)}$$ (6)

where material constants are given by

$$Q_{11} = \frac{E_1}{1-v_{12}v_{21}}, Q_{12} = \frac{v_{12}E_2}{1-v_{12}v_{21}}, Q_{22} = \frac{E_2}{1-v_{12}v_{21}}$$
$$Q_{66} = G_{12}, Q_{55} = G_{13}, Q_{44} = G_{23}$$ (7)

in which $E_1$, $E_2$ are the Young modulus in the 1 and 2 directions, respectively, and $G_{12}$, $G_{23}$, $G_{13}$ are the shear modulus in the 1-2, 2-3, 3-1 planes, respectively, and $v_{ij}$ are Poisson's ratios.

The laminate is usually made of several orthotropic layers in which the stress-strain relation for the $k^{th}$ orthotropic lamina with the arbitrary fiber orientation compared to the reference axes is given by [3]

$$\begin{Bmatrix} \sigma_{xx} \\ \sigma_{yy} \\ \tau_{xy} \\ \tau_{xz} \\ \tau_{yz} \end{Bmatrix}^{(k)} = \begin{bmatrix} \bar{Q}_{11} & \bar{Q}_{12} & \bar{Q}_{16} & 0 & 0 \\ \bar{Q}_{21} & \bar{Q}_{22} & \bar{Q}_{26} & 0 & 0 \\ \bar{Q}_{61} & \bar{Q}_{62} & \bar{Q}_{66} & 0 & 0 \\ 0 & 0 & 0 & \bar{Q}_{55} & \bar{Q}_{54} \\ 0 & 0 & 0 & \bar{Q}_{45} & \bar{Q}_{44} \end{bmatrix}^{(k)} \begin{Bmatrix} \varepsilon_{xx} \\ \varepsilon_{yy} \\ \gamma_{xy} \\ \gamma_{xz} \\ \gamma_{yz} \end{Bmatrix}^{(k)}$$ (8)

The inplane force, moments and shear force are defined as

$$\begin{Bmatrix} N_{ij} \\ M_{ij} \\ P_{ij} \end{Bmatrix} = \int_{-h/2}^{h/2} \sigma_{ij} \begin{Bmatrix} 1 \\ z \\ f(z) \end{Bmatrix} dz \quad \text{and} \quad \begin{Bmatrix} Q_x \\ Q_y \end{Bmatrix} = \int_{-h/2}^{h/2} f'(z) \begin{Bmatrix} \tau_{xz} \\ \tau_{yz} \end{Bmatrix} dz \tag{9}$$

It is noted that the function $f'(z) = 1 - 4z^2/h^2$ gets zero values at $z = \pm h/2$. It means that the traction-free boundary condition is automatically satisfied at the top and bottom plate surfaces. Furthermore, the transverse shear forces are described parabolically through the plate thickness. Hence, the shear correction factors are not required in this model.

Substituting Eq. (8) into Eq. (9), stress resultants are rewritten by

$$\hat{\boldsymbol{\sigma}} = \begin{Bmatrix} \mathbf{N} \\ \mathbf{M} \\ \mathbf{P} \\ \mathbf{Q} \end{Bmatrix} = \begin{bmatrix} \mathbf{A} & \mathbf{B} & \mathbf{E} & \mathbf{0} \\ \mathbf{B} & \mathbf{D} & \mathbf{F} & \mathbf{0} \\ \mathbf{E} & \mathbf{F} & \mathbf{H} & \mathbf{0} \\ \mathbf{0} & \mathbf{0} & \mathbf{0} & \mathbf{D}^S \end{bmatrix} \begin{Bmatrix} \boldsymbol{\varepsilon}_m \\ \boldsymbol{\kappa}_1 \\ \boldsymbol{\kappa}_2 \\ \boldsymbol{\beta} \end{Bmatrix} = \hat{\mathbf{D}} \hat{\boldsymbol{\varepsilon}} \tag{10}$$

where

$$A_{ij}, B_{ij}, D_{ij}, E_{ij}, F_{ij}, H_{ij} = \int_{-h/2}^{h/2} \left(1, z, z^2, f(z), zf(z), f^2(z)\right) \bar{Q}_{ij} dz , \quad i,j = 1,2,6$$

$$D_{ij}^S = \int_{-h/2}^{h/2} [f'(z)]^2 \bar{Q}_{ij} dz , \quad i,j = 4,5 \tag{11}$$

and the generalized strain $\hat{\boldsymbol{\varepsilon}}$ is divided into the linear and nonlinear strain components $\hat{\boldsymbol{\varepsilon}}_L = [\boldsymbol{\varepsilon}_L \ \boldsymbol{\kappa}_1 \ \boldsymbol{\kappa}_2 \ \boldsymbol{\beta}]^T$ and $\hat{\boldsymbol{\varepsilon}}_{NL} = [\boldsymbol{\varepsilon}_{NL} \ 0 \ 0 \ 0]^T$, respectively

$$\hat{\boldsymbol{\varepsilon}} = \hat{\boldsymbol{\varepsilon}}_L + \hat{\boldsymbol{\varepsilon}}_{NL} \tag{12}$$

Neglecting damping effect, the equation of motion obtained from Lagrange's equation using Hamilton's variation principle can be briefly expressed as [28]

$$\int_\Omega \hat{\boldsymbol{\sigma}}^T \delta \hat{\boldsymbol{\varepsilon}} d\Omega + \int_\Omega \delta \tilde{\mathbf{u}}^T \mathbf{m} \ddot{\tilde{\mathbf{u}}} d\Omega = \int_\Omega \mathbf{f}_s \delta \mathbf{u}^T d\Omega \tag{13}$$

where $\mathbf{m}$ and $\mathbf{f}_s$ are the consistent mass matrix (detailed in [8]) and the mechanical surface loads, respectively and

$$\tilde{\mathbf{u}} = \{\mathbf{u}_1 \ \mathbf{u}_2 \ \mathbf{u}_3\}^T \tag{14}$$

## 3. Isogeometric formulation for nonlinear analysis of plate

*3.1. A brief of NURBS basic functions*

A knot vector $\Xi = \{\xi_1, \xi_2, ..., \xi_{n+p+1}\}$ is a non-decreasing sequence of parameter values $\xi_i$, $i = 1,...n+p$, where $\xi_i \in R$ called $i^{th}$ knot lies in the parametric space, $p$ is the order of the B-spline and $n$ is number of basis functions. In the so-called open knot vector, the first and the last knots are repeated $p+1$ times and have the value of 0 and 1, respectively.

Using Cox-de Boor algorithm, the univariate B-spline basis functions $N_{i,p}(\xi)$ are defined recursively [29] on the corresponding knot vector start with order $p = 0$

$$N_{i,0}(\xi) = \begin{cases} 1 & \text{if } \xi_i \leq \xi < \xi_{i+1} \\ 0 & \text{otherwise} \end{cases} \tag{15}$$

as $p \geq 1$ the basis functions are obtained from

$$N_{i,p}(\xi) = \frac{\xi - \xi_i}{\xi_{i+p} - \xi_i} N_{i,p-1}(\xi) + \frac{\xi_{i+p+1} - \xi}{\xi_{i+p+1} - \xi_{i+1}} N_{i+1,p-1}(\xi) \tag{16}$$

An example of B-spline basis is illustrated in Figure 1. Using two open knot vectors $\Xi = \{0, 0, 0, \frac{1}{5}, \frac{2}{5}, \frac{3}{5}, \frac{3}{5}, \frac{4}{5}, 1, 1, 1\}$ and $\Xi = \{0, 0, 0, 0, \frac{1}{4}, \frac{1}{2}, \frac{3}{4}, 1, 1, 1, 1\}$, the two sets of univariate quadratic and cubic B-splines are plotted in Figure 1a and b, respectively. The important properties of B-Spline functions are summarized as below:

- Being piecewise polynomial functions.

- Non-negative over the domain $N_{i,p}(\xi) \geq 0$, $\forall \xi, i, p$

- Constitute partition of unity $\sum N_{i,p}(\xi) = 1$, $\forall \xi$

- $C^{p-k}$ continuity at the interior knot with $k$ is repeated time of this knot.

By a simple way – so-called tensor product of univariate B-splines, the multivariate B-spline basis functions are generated

$$N_A(\boldsymbol{\xi}) = \prod_{\alpha=1}^{d} N_{i_\alpha, p_\alpha}(\xi^\alpha) \qquad (17)$$

where parametric $d = 1, 2, 3$ according to 1D, 2D and 3D spaces, respectively. Figure 1c illustrates and example of bivariate B-spline basis functions according to two mentioned univariate B-splines.

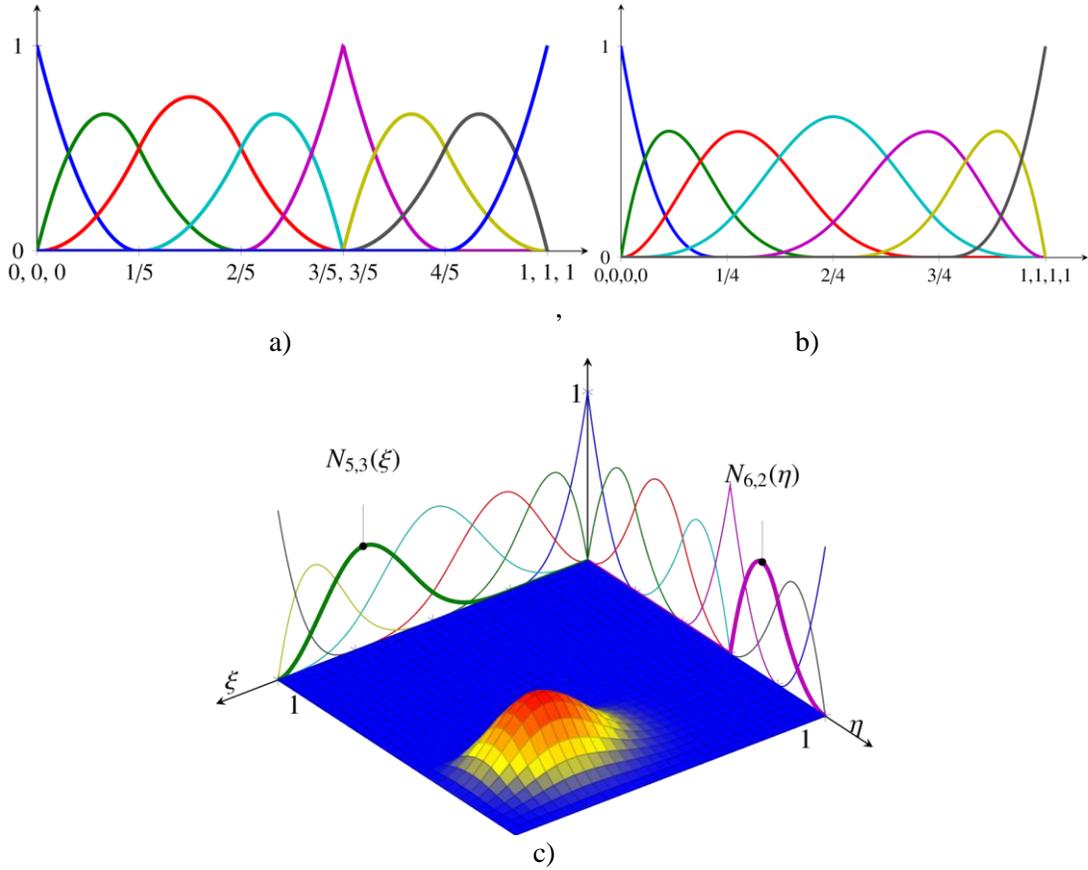

Figure 1. B-splines basic functions: a) Univariate quadratic; b) Univariate cubic and c) Bivariate.

To present exactly some conic sections, e.g., circles, cylinders, spheres, etc., non-uniform rational B-splines (NURBS) need to be used. Being different from B-spline, each control point of NURBS has an additional value called an individual weight $\zeta_A > 0$.

$$R_A(\xi,\eta) = \frac{N_A \zeta_A}{\sum_{A}^{m \times n} N_A(\xi,\eta)\zeta_A} \qquad (18)$$

NURBS basic functions also inherit all of features of B-spline basic functions and become B-spline basic functions when the individual weight of control points is constant.

*3.2. NURBS formulation for nonlinear bending of composite plates*

Using NURBS basis functions, the displacement field of the plate is approximated as

$$\mathbf{u}^h(\xi,\eta) = \sum_{A}^{m \times n} R_A(\xi,\eta) \mathbf{q}_A \qquad (19)$$

where $\mathbf{q}_A = \begin{bmatrix} u_{0A} & v_{0A} & \beta_{xA} & \beta_{yA} & w_A \end{bmatrix}^T$ is the vector of nodal degrees of freedom associated with the control point $A$.

Substituting Eq. (19) into Eq. (12), the generalized strains can be rewritten as:

$$\hat{\boldsymbol{\varepsilon}} = \sum_{A=1}^{m \times n} \left( \mathbf{B}_A^L + \frac{1}{2} \mathbf{B}_A^{NL} \right) \mathbf{q}_A \qquad (20)$$

where

$$\mathbf{B}_A^L = \left[ \left(\mathbf{B}_A^m\right)^T \ \left(\mathbf{B}_A^{b1}\right)^T \ \left(\mathbf{B}_A^{b2}\right)^T \ \left(\mathbf{B}_A^s\right)^T \right]^T$$

in which

$$\mathbf{B}_A^m = \begin{bmatrix} R_{A,x} & 0 & 0 & 0 & 0 \\ 0 & R_{A,y} & 0 & 0 & 0 \\ R_{A,y} & R_{A,x} & 0 & 0 & 0 \end{bmatrix}, \quad \mathbf{B}_A^{b1} = -\begin{bmatrix} 0 & 0 & R_{A,xx} & 0 & 0 \\ 0 & 0 & R_{A,yy} & 0 & 0 \\ 0 & 0 & 2R_{A,xy} & 0 & 0 \end{bmatrix}, \qquad (21)$$

$$\mathbf{B}_A^{b2} = \begin{bmatrix} 0 & 0 & 0 & R_{A,x} & 0 \\ 0 & 0 & 0 & 0 & R_{A,y} \\ 0 & 0 & 0 & R_{A,y} & R_{A,x} \end{bmatrix}, \quad \mathbf{B}_A^s = \begin{bmatrix} 0 & 0 & 0 & R_A & 0 \\ 0 & 0 & 0 & 0 & R_A \end{bmatrix}$$

while strain matrix $\mathbf{B}_A^{NL}$ is still dependent upon displacement gradient

$$\mathbf{B}_A^{NL}(\mathbf{q}) = \begin{bmatrix} \mathbf{A}_\theta \\ \mathbf{0} \end{bmatrix} \mathbf{B}_A^g \quad \text{where} \quad \mathbf{B}_A^g = \begin{bmatrix} 0 & 0 & R_{A,x} & 0 & 0 \\ 0 & 0 & R_{A,y} & 0 & 0 \end{bmatrix} \tag{22}$$

Substituting Eq. (20) into Eq. (13), and after eliminating the virtual displacement, the equation of motion is written in the following matrix form

$$\mathbf{Kq} + \mathbf{M\ddot{q}} = \mathbf{F}^{ext} \tag{23}$$

where **K** and **M** are the global stiffness and mass matrices, respectively

$$\mathbf{K}(\mathbf{q}) = \int_\Omega \left(\mathbf{B}^L + \mathbf{B}^{NL}\right)^T \hat{\mathbf{D}} \left(\mathbf{B}^L + 0.5\mathbf{B}^{NL}\right) d\Omega \tag{24}$$

$$\mathbf{M} = \int_\Omega \tilde{\mathbf{N}}^T \mathbf{m} \tilde{\mathbf{N}} d\Omega \tag{25}$$

in which

$$\tilde{\mathbf{N}}_A = \begin{Bmatrix} \mathbf{N}_A^1 \\ \mathbf{N}_A^2 \\ \mathbf{N}_A^3 \end{Bmatrix}, \qquad \mathbf{N}_A^1 = \begin{bmatrix} R_A & 0 & 0 & 0 & 0 \\ 0 & R_A & 0 & 0 & 0 \\ 0 & 0 & R_A & 0 & 0 \end{bmatrix};$$

$$\mathbf{N}_A^2 = -\begin{bmatrix} 0 & 0 & R_{A,x} & 0 & 0 \\ 0 & 0 & R_{A,y} & 0 & 0 \\ 0 & 0 & 0 & 0 & 0 \end{bmatrix}; \quad \mathbf{N}_A^3 = \begin{bmatrix} 0 & 0 & 0 & R_A & 0 \\ 0 & 0 & 0 & 0 & R_A \\ 0 & 0 & 0 & 0 & 0 \end{bmatrix} \tag{26}$$

and the external force vector under the transverse load $f_0$ is computed by

$$\mathbf{F}_A^{ext} = \int_\Omega f_0(t) \begin{bmatrix} 0 & 0 & R_A & 0 & 0 \end{bmatrix}^T d\Omega \tag{27}$$

## 4. Solution scheme

From Eq. (23), it is observed that the equation of dynamic system is dependent upon both time domain and unknown displacement vector. To discretize this problem, the Newmark's integration scheme association with the iteration methods is employed.

### 4.1. Time integration

The dynamic problem is solved in step-by-step of a number of equal time intervals, $\Delta t$ with zero displacement, velocity and acceleration at initial time, $t = 0$. And the first and second derivative of displacement are sought implicitly at time $(m+1)\Delta t$ as below

$$\ddot{\mathbf{q}}_{m+1} = \frac{1}{\beta \Delta t^2}(\mathbf{q}_{m+1} - \mathbf{q}_m) - \frac{1}{\beta \Delta t}\dot{\mathbf{q}}_m - \left(\frac{1}{2\beta} - 1\right)\ddot{\mathbf{q}}_m \qquad (28)$$

$$\dot{\mathbf{q}}_{m+1} = \dot{\mathbf{q}}_m + \Delta t(1-\gamma)\ddot{\mathbf{q}}_m + \gamma \Delta t \ddot{\mathbf{q}}_{m+1} \qquad (29)$$

where $\beta$ and $\gamma$ are constant and equal to 0.25 and 0.5, respectively [30].

Substituting Eq. (28) into Eq. (23), we obtain

$$\hat{\mathbf{K}}_{m+1}\mathbf{q}_{m+1} = \hat{\mathbf{F}}_{m+1} \qquad (30)$$

where $\hat{\mathbf{K}}_{m+1}$ and $\hat{\mathbf{F}}_{m+1}$ are the effective stiffness matrix and force vector at time $(m+1)\Delta t$

$$\hat{\mathbf{K}}_{m+1} = \mathbf{K}_{m+1} + \frac{1}{\beta \Delta t^2}\mathbf{M},$$

$$\hat{\mathbf{F}}_{m+1} = \mathbf{F}_{m+1}^{ext} + \mathbf{M}\left[\frac{1}{\beta \Delta t^2}\mathbf{q}_m + \frac{1}{\beta \Delta t}\dot{\mathbf{q}}_m + \left(\frac{1}{2\beta} - 1\right)\ddot{\mathbf{q}}_m\right] \qquad (31)$$

Referring to Eq.(30), the right hand side is known from the converged solutions at previous time step, i.e. $t = m\Delta t$. Since, the effective stiffness $\hat{\mathbf{K}}_{m+1}$ is nonlinearly dependent on the displacements $\mathbf{q}_{m+1}$, this equation must be computed as an iterative processing such as the Picard algorithm or the Newton-Raphson method which can be presented in the next section.

*4.2. Iteration methods*

At time step $(m+1)\Delta t$, Eq.(30) can be rewritten in term of the residual force as below

$$\boldsymbol{\varphi}_{m+1} = \hat{\mathbf{K}}_{m+1}\mathbf{q}_{m+1} - \hat{\mathbf{F}}_{m+1} \qquad (32)$$

The residual force presents the error in this approximation and tends to zero during iteration. If $^i\mathbf{q}_{m+1}$, an approximate trial solution at the $i^{th}$ iteration, makes unbalance residual force, an improved solution is then proposed

$$^{i+1}\mathbf{q}_{m+1} = {}^i\mathbf{q}_{m+1} + \Delta \mathbf{q} \qquad (33)$$

where the incremental displacement is calculated by equaling to zero curtailed Taylor's series expansion of $^{i+1}\boldsymbol{\varphi}_{m+1}$ [31]

$$\Delta \mathbf{q} = -{}^i\boldsymbol{\varphi}_{m+1} / \mathbf{K}_T \qquad (34)$$

in which $\mathbf{K}_T$ is called tangent stiffness matrix. If it takes the same form of the effective stiffness matrix given in Eq.(31), this iterative method is so-called the direct iteration method or Picard method. This method is simple in concept and implementation. But sometimes, it does not work because it is too hard to get invertible form of the unsymmetric stiffness matrix. Another way, $\mathbf{K}_T$ could be computed following to the Newton-Raphson method

$$\mathbf{K}_T = \partial \boldsymbol{\varphi}(^i\mathbf{q}) / \partial \mathbf{q} \tag{35}$$

It is known that this matrix $\mathbf{K}_T$ is always symmetric for all structure problems and it helps this method converges faster for most applications than the Picard one [32].

At each time step, the process in Eq.(33) is repeated until the displacement error between two consecutive iterations reduces to the desired error tolerance.

$$\frac{\left\| ^{i+1}\mathbf{q}_{m+1} - {^i}\mathbf{q}_{m+1} \right\|}{\left\| ^i\mathbf{q}_{m+1} \right\|} < tol \tag{36}$$

## 5. Numerical results

In this section, we show performance of IGA for several geometrically nonlinear plate problems using HSDT. In previous works [8, 9, 21], it is found that just with $11 \times 11$ cubic NURBS elements, IGA using HSDT produces an ultra-accurate solutions for plate analysis. In this study, this mesh therefore is employed.

Some sets of the material properties are used for numerical investigations:

Material *I*:

$E_1 = 3 \times 10^6 \, \text{psi}, \; E_2 = 1.28 \times 10^6 \, \text{psi}, \; G_{12} = G_{23} = G_{31} = 0.37 \times 10^6 \, \text{psi}, \; \nu = 0.32$

Material *II*

$E_1 = 1.8282 \times 10^6 \, \text{psi}, \; E_2 = 1.8315 \times 10^6 \, \text{psi}, \; G_{12} = G_{23} = G_{31} = 0.3125 \times 10^6 \, \text{psi}, \; \nu = 0.2395$

Material *III*

$E_1 = 25 E_2, \; G_{12} = G_{13} = 0.5 E_2, \; G_{23} = 0.2 E_2, \; \nu_{12} = 0.25.$

Material *IV*

$E_1 = 40 E_2, \; G_{12} = G_{13} = 0.6 E_2, \; G_{23} = 0.5 E_2, \; \nu_{12} = 0.25.$

Material *V* [33]

$E_1 = 525$ GPa, $E_2 = 21$ GPa, $G_{12} = G_{23} = G_{13} = 10.5$ GPa, $v_{12} = 0.25$, $\rho = 800$ kg/m$^3$.

Material *VI* [34]

$E_1 = 172.369$ GPa, $E_2 = 6.895$ GPa, $G_{12} = G_{13} = 3.448$ GPa, $G_{23} = 1.379$ GPa, $v_{12} = 0.25$, $\rho = 1603.03$ kg/m$^3$.

In all examples, the convergence tolerance is taken to be 1%. For convenience, the non-dimensional load parameter, transverse displacement, stresses are expressed below, unless specified otherwise

$$\bar{P} = \frac{q_0 a^4}{E_2 h^3}, \quad \bar{w} = \frac{w}{h}, \quad \bar{\sigma} = \frac{\sigma L^2}{E h^2}$$

*5.1. Geometrically nonlinear bending analysis*

In this section, the geometrically nonlinear bending analysis of plate is studied by computing Eq. (23) without the mass matrix effect

$$\mathbf{Kq} = \mathbf{F}^{ext} \tag{37}$$

Herein, the global stiffness matrix $\mathbf{K}$ is still nonlinear relation with unknown variable $\mathbf{q}$ because of dependence on nonlinear strain matrix $\mathbf{B}^{NL}$. To solve this equation, Newton-Raphson methods is employed to calculate the tangent stiffness as follow:

$$\mathbf{K}_T = \mathbf{K}_L + \mathbf{K}_{NL} + \mathbf{K}_g \tag{38}$$

where $\mathbf{K}_L$, $\mathbf{K}_{NL}$, $\mathbf{K}_g$ are the linear, nonlinear and geometric stiffness matrix, respectively

$$\begin{aligned}
\mathbf{K}_L &= \int_\Omega \left(\mathbf{B}^L\right)^T \hat{\mathbf{D}} \mathbf{B}^L d\Omega \\
\mathbf{K}_{NL} &= \int_\Omega \left(\mathbf{B}^L\right)^T \hat{\mathbf{D}} \mathbf{B}^{NL} + \left(\mathbf{B}^{NL}\right)^T \hat{\mathbf{D}} \mathbf{B}^N + \left(\mathbf{B}^{NL}\right)^T \hat{\mathbf{D}} \mathbf{B}^{NL} d\Omega \\
\mathbf{K}_g &= \int_\Omega \left(\mathbf{B}^g\right)^T \mathbf{N}_0 \mathbf{B}^g d\Omega
\end{aligned} \tag{39}$$

in which $\mathbf{N}_0 = \begin{bmatrix} N_x^0 & N_{xy}^0 \\ N_{xy}^0 & N_y^0 \end{bmatrix}$ is a matrix related to the in-plane forces.

*5.1.1. Isotropic plates*

Let us consider a clamped square thin plate under a uniformly distributed load. This is a benchmark problem which is often tested by many researchers [31, 35, 36] for the geometrically nonlinear validation of thin plate formulation. Table 1 provides the results for central deflection $\bar{w}$ and axial stress $\bar{\sigma}_x(0,0,h/2)$. The obtained solutions are compared with the analytical ones using a double Fourier series by Levy [35] and finite element ones using nine node element [31], and mixed finite element solution [36]. Their relations with load parameter are also depicted in Figure 2. As seen, the present results are in excellent agreement with those from the literature and gain the best axial stress as compared to analytical Kirchhoff solution.

Table 1: Central deflection and axial stress of a clamped square plate under uniform load

| Load | Anal. Sol. [35] | | MXFEM[36] | | FEM Q9 [31] | | IGA | |
|---|---|---|---|---|---|---|---|---|
| $\bar{P}$ | $\bar{w}$ | $\bar{\sigma}_x$ | $\bar{w}$ | $\bar{\sigma}_x$ | $\bar{w}$ | $\bar{\sigma}_x$ | $\bar{w}$ | $\bar{\sigma}_x$ |
| 17.8 | 0.237 | 2.6 | 0.2392 | 2.414 | 0.2361 | 2.614 | 0.2367 | 2.5626 |
| 38.3 | 0.471 | 5.2 | 0.4738 | 5.022 | 0.4687 | 5.452 | 0.4693 | 5.3273 |
| 63.4 | 0.695 | 8.0 | 0.6965 | 7.649 | 0.6902 | 8.291 | 0.6910 | 8.0998 |
| 95.0 | 0.912 | 11.1 | 0.9087 | 10.254 | 0.9015 | 11.066 | 0.9025 | 10.8273 |
| 134.9 | 1.121 | 13.3 | 1.1130 | 12.850 | 1.1050 | 13.789 | 1.1061 | 13.5223 |
| 184.0 | 1.323 | 15.9 | 1.3080 | 15.420 | 1.2997 | 16.456 | 1.3009 | 16.1806 |
| 245.0 | 1.521 | 19.2 | 1.5010 | 18.060 | 1.4916 | 19.178 | 1.4928 | 18.9069 |
| 318.0 | 1.714 | 21.9 | 1.6880 | 20.741 | 1.6775 | 21.938 | 1.6786 | 21.6797 |
| 402.0 | 1.902 | 25.1 | 1.8660 | 23.423 | 1.8545 | 24.713 | 1.8555 | 24.4700 |

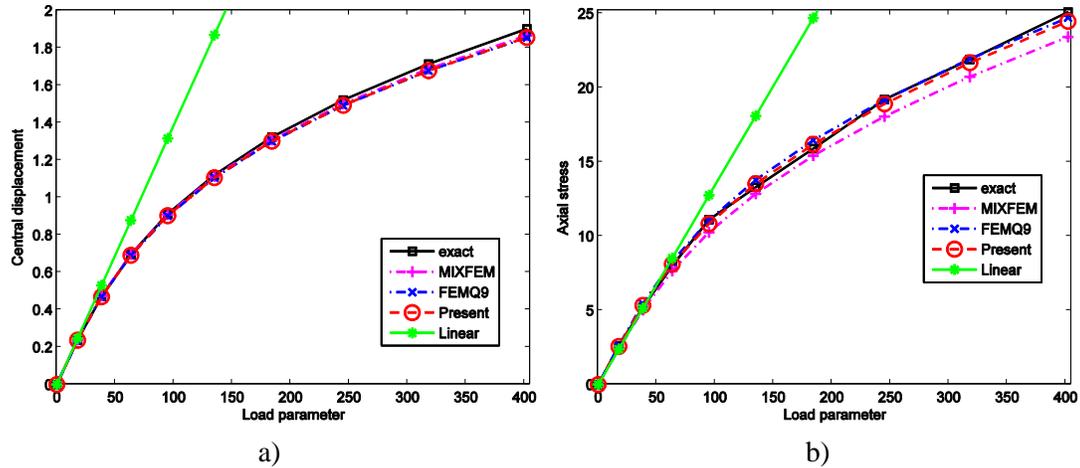

a)                                                          b)

Figure 2. Comparison of the non-dimensional central deflection (a) and normal stress (b) of isotropic plate.

Next, the large deformation analysis of a circular plate under uniform pressure is considered. We consider the plate with geometric data: radius to thickness ratio $R/h = 50$ and material properties: Young's modulus $E = 10^7$ and Poisson's ratio $\nu = 0.3$. The present transverse central displacements listed in Table 2 are compared with the analytical Kirchhoff solution [37], that of Kirchhoff-based elements such as DKT[38], RNEM [39] and that of Mindlin-based elements: nine-node Lagrangian quadrilateral element (QL) [31], mixed interpolation smoothing quadrilateral element with 20DOF (MISQ20) [40]. As seen, the present method produces the most accurate solution.

Table 2: Normalized central deflection of a clamped circular plate under uniform load

| Load $\bar{P}$ | Normalized central deflection $\bar{w}$ | | | | | Anal. Sol. [37] |
|---|---|---|---|---|---|---|
| | MISQ20 [40] | QL [31] | DKT[38] | RNEM [39] | IGA | |
| 1 | 0.170 (0.59) | 0.1682 (0.47) | 0.172 (1.78) | 0.1664 (1.54) | 0.1669 (1.24) | 0.169 |
| 2 | 0.327 (1.24) | 0.3231 (0.03) | 0.330 (2.17) | 0.3179 (1.58) | 0.3208 (0.68) | 0.323 |
| 3 | 0.465 (1.75) | 0.4591 (0.46) | 0.470 (2.84) | 0.4514 (1.23) | 0.4562 (0.18) | 0.457 |
| 6 | 0.780 (2.50) | 0.7702 (1.21) | 0.791 (3.94) | 0.7637 (0.35) | 0.7671 (0.80) | 0.761 |
| 10 | 1.067 (3.09) | 1.0514 (1.58) | 1.082 (4.54) | 1.0544 (1.87) | 1.0487 (1.32) | 1.035 |
| 15 | 1.320 (3.21) | 1.3007 (1.70) | 1.342 (4.93) | 1.3164 (2.92) | 1.2989 (1.56) | 1.279 |

The errors compared to analytical Kirchhoff solution [37] are shown in parenthesis.

*5.1.2. Symmetric laminated plates*

Firstly, the benchmark problems with the experimental results of Zaghloul and Kenedy are studied for validating the present method. Figure 3 reveals comparison between the present solutions based on the HSDT model with others according to CPT [41], FSDT [36] and the experimental results [41]. Figure 3a shows the relation between the central deflection and the uniform load intensity for an orthotropic plate (material set *I*) under simply supported constraint (SSSS)

$$u_0 = w = \beta_x = 0 \quad \text{at lower and upper edges}$$
$$v_0 = w = \beta_y = 0 \quad \text{at left and right edges}$$
(40)

It is clearly seen that considering the shear deformation effect, the IGA results with 5 DOFs/control points match well with those of FSDT [36] using MIXFEM with 8

DOFs/control points and get good agreement with experimental ones. The same conclusion is observed for a clamped 4-crosply [0/90/90/0] plate (material set *II*) which plotted in Figure 3b

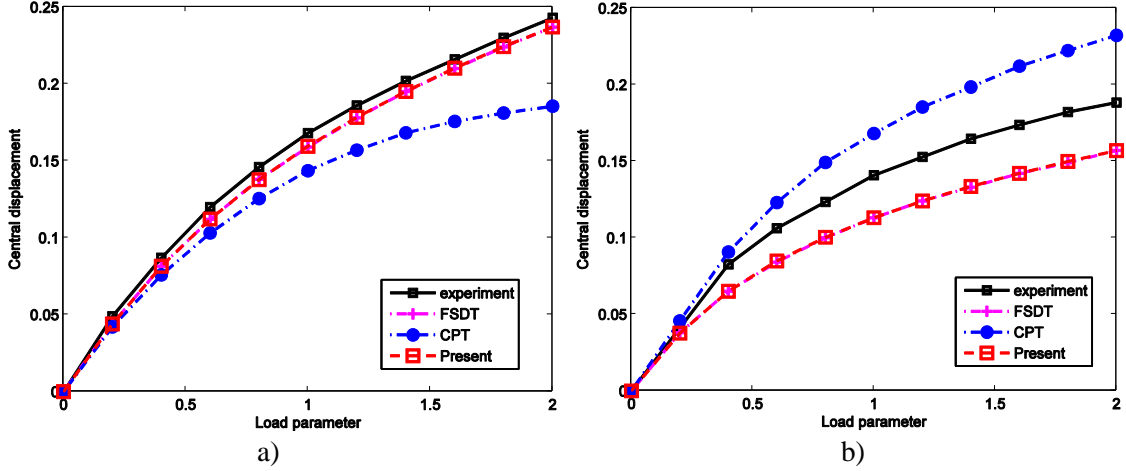

Figure 3. The load-deflection curves of: (a) the simply supported plate
($L=12$in, $h=0.138$in) and (b) the clamped plate ($L=12$in, $h=0.096$in)[†].

Next, the effect of span to thickness ratio on the central deflection of the symmetric laminated composite plates is revealed in Table 3 and Table 4. Herein, material *III* is set. In Table 3 the central deflection versus the load parameter is depicted for the plate subjected to uniform load and simply supported constraint (SSSS2)

$$u_0 = v_0 = w = \beta_x = 0 \quad \text{at lower and upper edges}$$
$$u_0 = v_0 = w = \beta_y = 0 \quad \text{at left and right edges}$$
(41)

The obtained results are compared with those of nine-node Lagrangian quadrilateral element with 9DOF based on $C^0$ HSDT [42], mesh free method based on SSDT [43]; and that of the FSDT model using MISQ20 [40], four-node quadrilateral isoparametric plane element with 20 and 24 DOFs [44]. It is interesting to note that just using 5 DOFs, present method gains very good agreement with other published solutions. Figure 4 shows that the present results match very well with the solutions by Kant [42] for various span to thickness ratios $L/h$ = 10, 20, 40. In Table 4, the linear and geometrically nonlinear solutions for simply supported (SSSS) of the [0/90/90/0] and [0/90/0] laminated plates are presented. In case of linear problem, Pagano has given the exact solutions by using 3D elasticity model [1]. However, the nonlinear one is not available. For a comparison purpose, the MITC element [45] is, thus, also used to compute the displacement for the plates based on the FSDT with SCF equals to 5/6. It can be seen that the HSDT model gets higher results than the FSDT with better accuracy as compare with Pagano's results. And the discrepancy between them increases according to increase in the thickness to length ratio or the magnitude of applied load. The clearer observation is found in Figure 5. Furthermore, in thick plate the geometrical nonlinearity is pronouncedly observed with more highly curved load-displacement line than that of the thin plate. Figure 6 plots the stress distributions through the plate thickness of the four cross-ply plate ($L/h$=10) via the change of load intensity. Regarding the nonlinear part in in-plane strain, the axial stress $\hat{\sigma} = \sigma h^2 / (q_0 L^2)$ unsymmetrically distributes through the mid-plane. And its magnitude at the bottom reduces faster than that at the top surface according to increase in the load intensity. Being different from the axial stress, shear stress $\hat{\tau}_{yz} = \tau_{yz} h / (q_0 a)$ is symmetric distribution and reduces according to increase in load parameter. Furthermore, the HSDT model enables to obtain the parabolic shape of the shear stresses, which naturally satisfy the traction-free boundary condition at the plate surfaces. To close this subsection, effect of the boundary conditions on the nonlinear behavior of [0/90/90/0] laminated plate is investigated. As expected, both normalized deflection $\hat{w} = 100 w E_2 h^3 / (q_0 L^4)$ and axial stress $\hat{\sigma}_x = \sigma_x h^2 / q_0 L^2$ reduce belong to change of boundary condition from SSSS to SSSS2 and CCCC because of increase in constraint at the boundary, which leads to stiffen the plate structure.

Table 3: Central deflection $\bar{w}$ of a simply supported (SSSS2) [0/90/90/0] square plate.

| $L/h$ | $\bar{P}$ | HOST [42] | MISQ20 [40] | RDKQ-NL20[44] | RDKQ-NL24 [44] | MQRBF [43] | Present |
|---|---|---|---|---|---|---|---|
| 40 | 50 | 0.293 | 0.296 | 0.291 | 0.294 | 0.2654 | 0.2936 |
|  | 100 | 0.464 | 0.473 | 0.461 | 0.467 | 0.446 | 0.4643 |
|  | 150 | 0.582 | 0.592 | 0.577 | 0.587 | 0.616 | 0.5798 |
|  | 200 | 0.664 | 0.683 | 0.667 | 0.679 | 0.7355 | 0.6683 |
|  | 250 | 0.738 | 0.759 | 0.74 | 0.754 | 0.8355 | 0.7407 |
| 20 | 50 | 0.320 | 0.312 | 0.323 | 0.327 | 0.3004 | 0.3126 |
|  | 100 | 0.493 | 0.487 | 0.487 | 0.494 | 0.5085 | 0.4807 |
|  | 150 | 0.592 | 0.603 | 0.597 | 0.608 | 0.6591 | 0.5928 |
|  | 200 | 0.680 | 0.691 | 0.682 | 0.695 | 0.778 | 0.6784 |
|  | 250 | 0.752 | 0.765 | 0.751 | 0.766 | 0.8771 | 0.7486 |

[†] The data is redrawn according to Urthaler's work [36].

| 10 | 50  | 0.360 | 0.356 | 0.363 | 0.37  | - | 0.3609 |
|    | 100 | 0.520 | 0.521 | 0.514 | 0.525 | - | 0.5179 |
|    | 150 | 0.624 | 0.629 | 0.616 | 0.629 | - | 0.6213 |
|    | 200 | 0.696 | 0.711 | 0.695 | 0.71  | - | 0.7005 |
|    | 250 | 0.760 | 0.779 | 0.761 | 0.777 | - | 0.7659 |

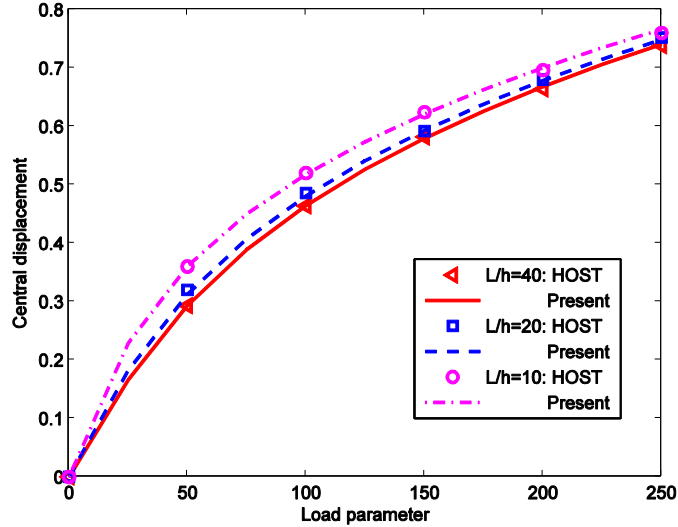

Figure 4. Comparison between $C^0$ HOST and present model for simply supported (SSSS2) [0/90/90/0] plate with various length to thickness ratios.

Table 4: Central deflection $\bar{w}$ of simply supported symmetric laminated composite plates

| L/h | $\bar{P}$ | [0/90/90/0] | | | | | [0/90/0] | | | | |
|---|---|---|---|---|---|---|---|---|---|---|---|
| | | Linear | | | Nonlinear | | Linear | | | Nonlinear | |
| | | HSDT | FSDT | 3D [1] | HSDT | FSDT | HSDT | FSDT | 3D [1] | HSDT | FSDT |
| 4 | 50  | 0.947 | 0.856 | 0.977 | 0.7198 | 0.6791 | 0.961 | 0.889 | 1.003 | 0.7262 | 0.6948 |
|   | 100 | 1.894 | 1.712 |       | 1.1214 | 1.0788 | 1.922 | 1.778 |       | 1.1284 | 1.0974 |
|   | 200 | 3.787 | 3.423 |       | 1.6555 | 1.6111 | 3.844 | 3.556 |       | 1.6606 | 1.6316 |
|   | 300 | 5.681 | 5.135 |       | 2.0447 | 1.9877 | 5.765 | 5.335 |       | 2.0475 | 2.0078 |
| 10 | 50  | 0.357 | 0.331 | 0.372 | 0.3474 | 0.3236 | 0.356 | 0.334 | 0.370 | 0.3462 | 0.3264 |
|    | 100 | 0.715 | 0.662 |       | 0.6501 | 0.6121 | 0.712 | 0.669 |       | 0.6478 | 0.6162 |
|    | 200 | 1.430 | 1.324 |       | 1.1148 | 1.0667 | 1.425 | 1.338 |       | 1.1116 | 1.0713 |
|    | 300 | 2.144 | 1.986 |       | 1.4612 | 1.4100 | 2.137 | 2.006 |       | 1.4586 | 1.4154 |
| 20 | 50  | 0.253 | 0.245 | 0.259 | 0.2504 | 0.2428 | 0.252 | 0.246 | -     | 0.2494 | 0.2432 |
|    | 100 | 0.506 | 0.490 |       | 0.4872 | 0.4734 | 0.504 | 0.491 |       | 0.4849 | 0.4737 |
|    | 200 | 1.012 | 0.980 |       | 0.8960 | 0.8763 | 1.008 | 0.982 |       | 0.8921 | 0.8752 |
|    | 300 | 1.518 | 1.470 |       | 1.2255 | 1.2024 | 1.513 | 1.473 |       | 1.2190 | 1.1999 |
| 100 | 50  | 0.217 | 0.216 | 0.217 | 0.2159 | 0.2150 | 0.217 | 0.216 | -     | 0.2158 | 0.2149 |
|     | 100 | 0.434 | 0.432 |       | 0.4243 | 0.4226 | 0.434 | 0.432 |       | 0.4238 | 0.4222 |
|     | 200 | 0.868 | 0.865 |       | 0.7993 | 0.7967 | 0.868 | 0.865 |       | 0.7969 | 0.7945 |
|     | 300 | 1.303 | 1.297 |       | 1.1146 | 1.1117 | 1.303 | 1.297 |       | 1.1101 | 1.1074 |

FSDT results using MITC element [45]

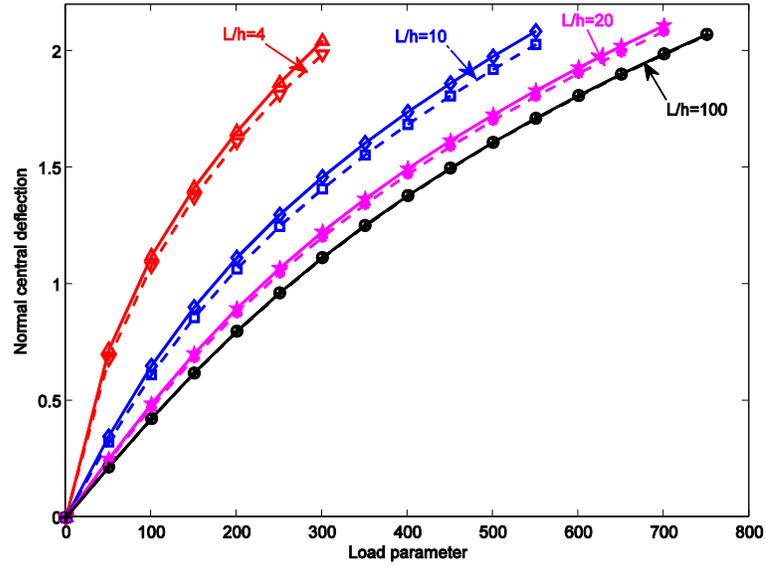

Figure 5. Load-displacement curve of symmetric laminated [0/90/90/0] plate: HSDT and FSDT results according to solid and dash lines, respectively. Their discrepancy increases by increasing in applied load or the length to thickness ratios.

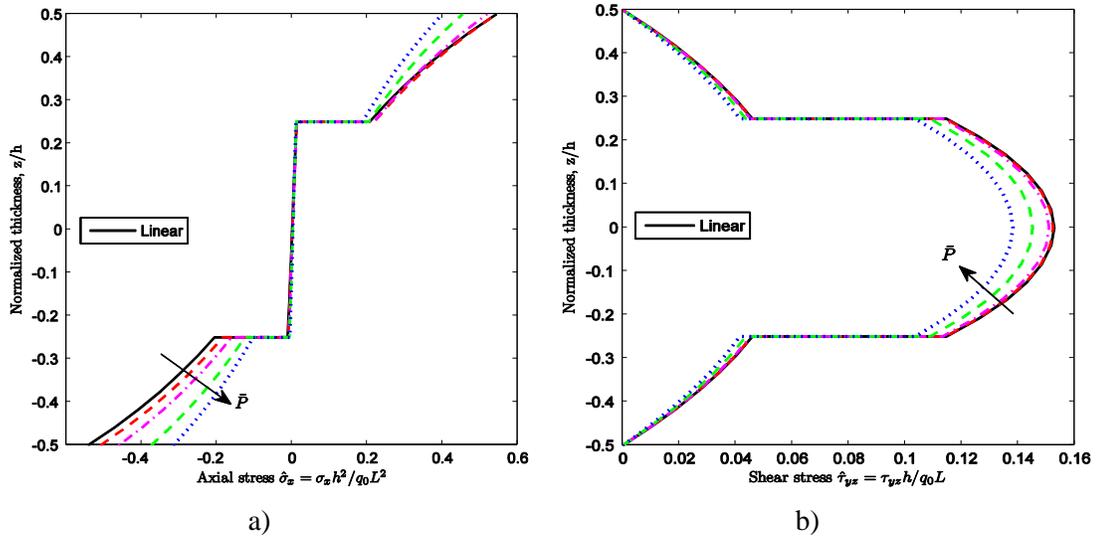

a)   b)

Figure 6. Effect of the load parameter $\bar{P}$ on the distributed stresses: the magnitude of load is 50, 100, 200 and 300 for the red, purple, green and blue curves, respectively.

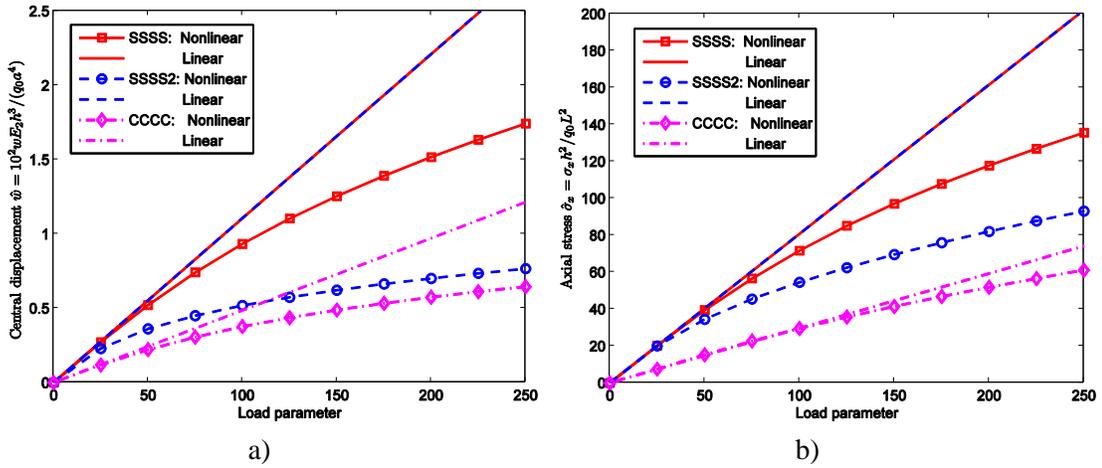

a)                  b)

Figure 7. Effect of the boundary conditions on the nonlinear behavior of [0/90/90/0] laminated plate under uniform pressure: (a) Central deflection and (b) Axial stress.

*5.1.3. Antiymmetric laminated plate*

This subsection deals with the analysis of the simply supported square laminated plate with $L/h$=10 and material *IV*. Figure 8 reveals the effect of number of layers on the nonlinear behavior of the $[0/90]_N$ cross-ply plates. It is observed that with the same total thickness, increase in number of layers *N* helps the plate stiffer with deflection reduction. Moreover, it also reduces the nonlinear effect on the laminated plate. It means that the load-deflection curve becomes closer to a straight line as the linear solution.

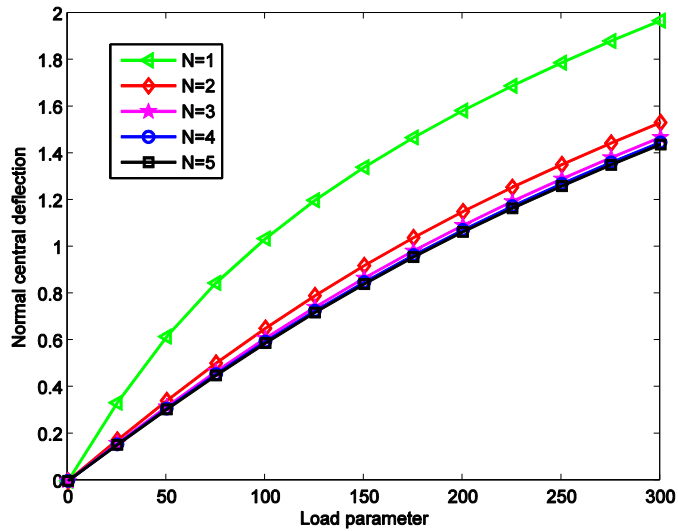

Figure 8. Effect of number of layers on deflection of the $[0/90]_N$ laminated composite plate.

Figure 9 shows the load – deflection curves of angle-ply [-$\theta$/$\theta$/-$\theta$/$\theta$] plate via the fiber orientation angle $\theta$ which changes from 0 to 15, 30, 45. As seen, the plate has the most stiffness at $\theta = 45°$. Figure 10 reveals the effect of fiber angle on the deflection behavior of both symmetric and anti-symmetric plates under various loading levels. The general observations are: (1) deflections are symmetric about angle $\theta = 45°$; (2) the plate behavior is the weakest in case of one orthotropic layer ($\theta = 0°$ or 90°) and the stiffest at $\theta = 45°$; (3) considering the bending-stretching coupling, the antisymmetric laminated plate obtains the lower deflection than the symmetric one and their discrepancy increases according to increase in angle from 0 to 45°; (4) nonlinear deflection parameter reduces following to increase in applied load and always lower than that of linear results.

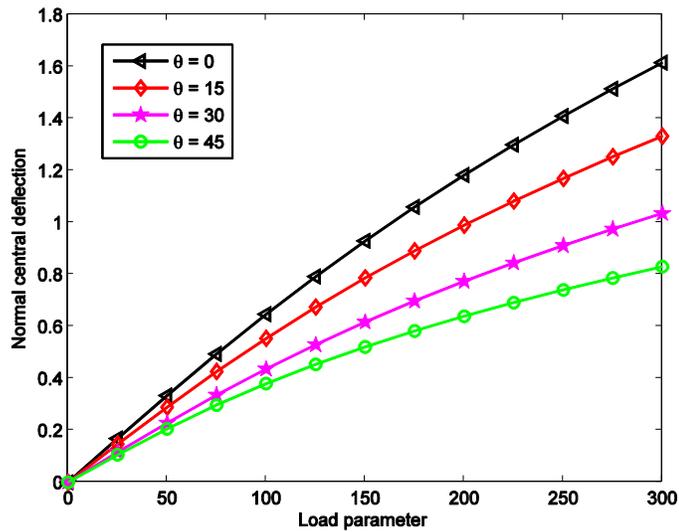

Figure 9. Effect of fiber orientation angle on load – deflection curves of the angle-ply [-$\theta$/$\theta$/-$\theta$/$\theta$] plate.

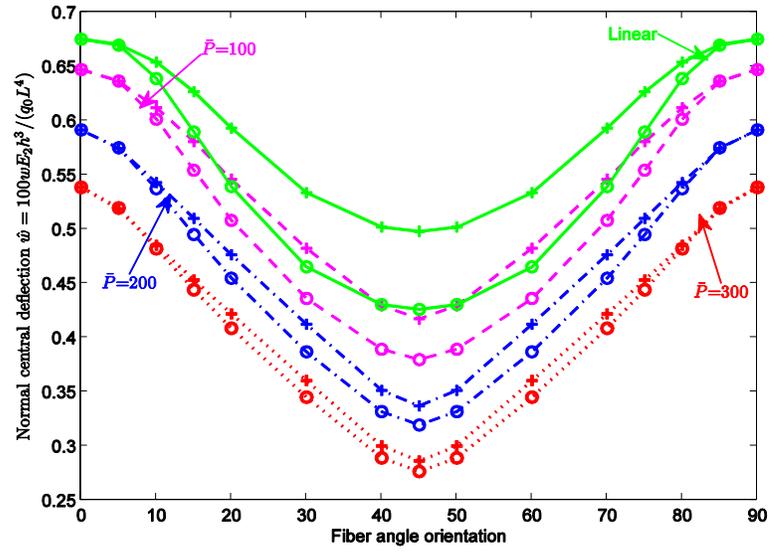

**+**: [-θ/θ/θ/-θ] symmetric laminated plate,   **O**: [-θ/θ/-θ/θ] antisymmetric laminated plate

Figure 10. Behavior of angle-ply plates under various loading levels.

*5.2. Geometrically nonlinear transient analysis*

In the geometrically nonlinear transient analysis, fully Lagrange equation motion in Eq. (23) is computed by a combined technique between Newmark's integration and Picard method. An orthotropic plate with set of material *V* and dimensions as length *L*=250 mm, thickness *h*=5 mm is firstly studied for validation. For this problem, the fully simply supported plate is subjected to a uniform step loading of 1 MPa. Its transient response according to the normalized central deflection $\bar{w}$ under both linear and nonlinear analysis is shown in Figure 11. It is observed that present method predicts the very close deflection response as compared with finite strip method (FSM) by J. Chen [33]. It also clearly exhibits that the magnitude and wavelength of the non-linear response are lower than that of linear behavior with the same loading intensity.

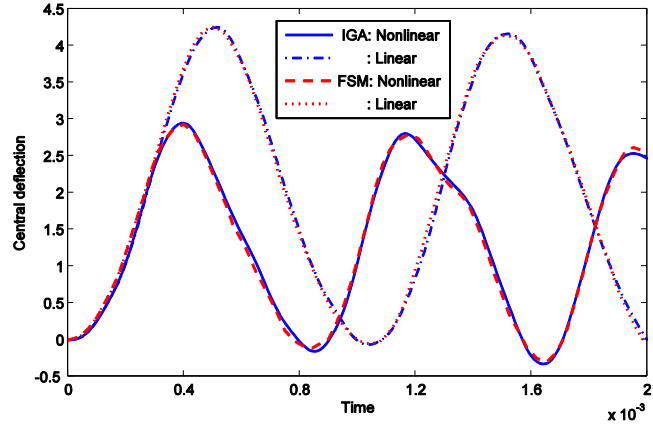

Figure 11. Time history of the transverse displacement of an orthotropic plate under step uniform load with intensity 1MPa.[‡]

Next, the dynamic response of three layer [0/90/0] thick plate is investigated. The material set *VI* is used for this plate ($L/h = 5$, $h = 0.1526$m). The transverse load is sinusoidally distributed in spatial domain and is assumed to vary with time as follows

$$f_0(x,y,t) = q_0 \sin(\frac{\pi x}{a})\sin(\frac{\pi y}{b})F_0(t) \tag{42}$$

in which $q_0 = 0.689$ GPa and value of force $F_0(t)$ depicted in Figure 12 depends on loading types: step, triangular, sinusoidal and explosive blast, respectively. Once again the observation is that nonlinear analysis takes the lower central deflection and higher frequency than that of the linear one.

---

[‡] The results by FSM is cited from [33].

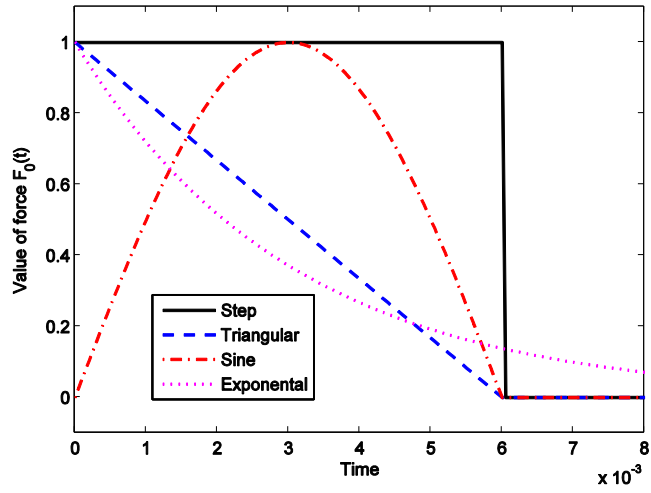

Figure 12. Time history of applied load $F_0(t)$.

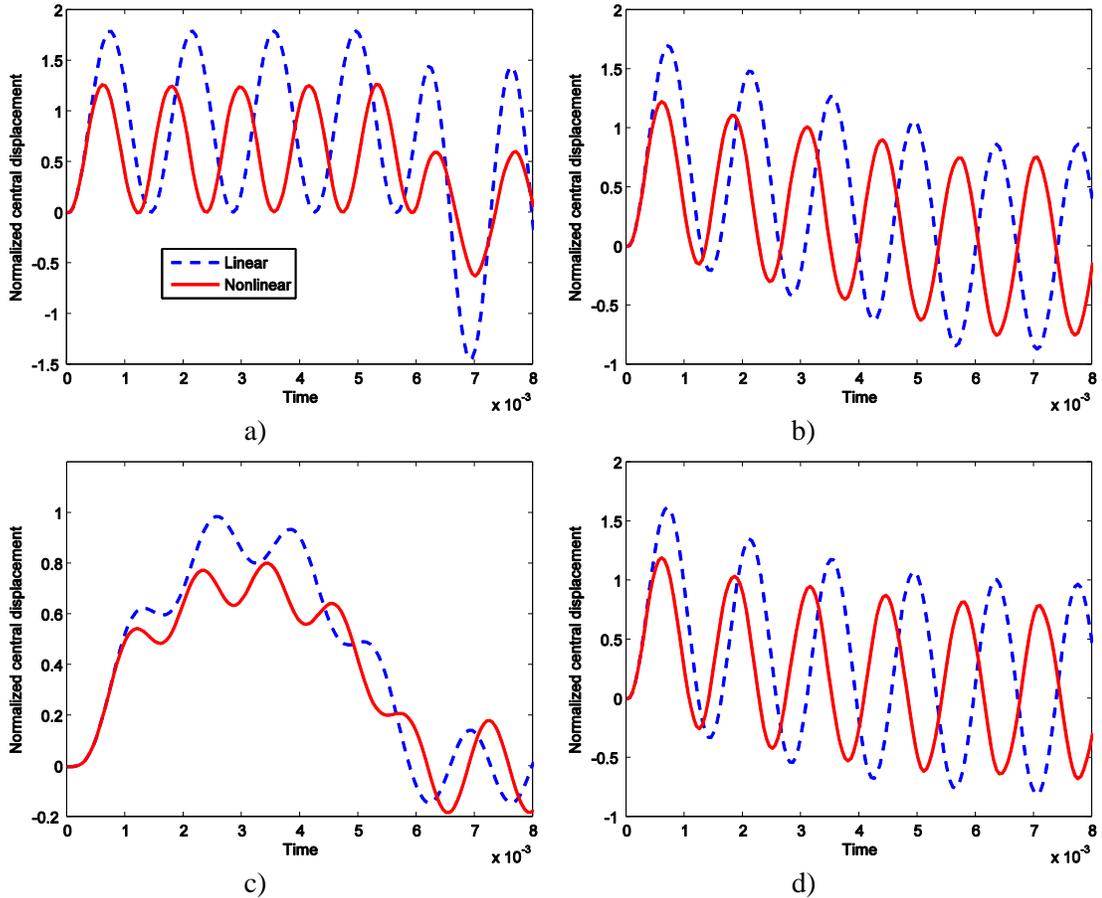

Figure 13. Effect of different loadings on the deflection respond of the cross-ply [0/90/0] square laminated plate: (a) step; (b) triangular; (c) sine and (d) explosive blast loading.

## 6. Conclusions

An effectively numerical procedure based on IGA and HSDT has been presented for geometrically nonlinear analysis of the laminated composite plates. Herein, using cubic approximation functions, the present method naturally satisfies the $C^1$ continuity across inter-element boundaries without any additional variables. The nonlinearity of the plates based on the von-Karman strain assumptions including nonlinear bending and transient problems which are solved by the Newmark time integration associated with the iteration methods. Numerous numerical examples have been carried out for isotropic, symmetric and unsymmetric laminated plates subjected to constant or dynamic loads. In all cases, the present results are in good agreement with available solutions in the literature. Furthermore, just utilizing five degrees of freedom, the present plate model remedies the shear locking phenomenon and shows very good performances for both thick and thin plates even though it is based on the thick plate theory. Especially, for clamped circular plate, present method gets the best results as compared with analytical Kirchhoff solution. It is believed that utilizing NURBS basic functions helps present method to eliminate the error of geometric approximation.

**Acknowledgements**

This research was supported by the Basic Science Research Program through the National Research Foundation of Korea (NRF) funded by the Ministry of Education, Science, and Technology (2010-019373 and 2012R1A2A1A01007405). The authors would like to acknowledge the Special Funding of Ghent University (Bijzonder Onderzoeksfonds), in the framework of BOF project BOF 01N02410. The third author appreciates for the support from the Vietnam National Foundation for Science and Technology Development (NAFOSTED) under grant number 107.02-2012.27.